\documentclass[11pt]{article}
\usepackage{amssymb,amsmath,epsfig}
\usepackage{cite}

\begin{document}

\title{\bf Collision of Particles near Charged MSW Black Hole in 2+1 Dimensions}

\author{\bf{Pameli Saha}\thanks{pameli.saha15@gmail.com}~ and
\bf{Ujjal Debnath}\thanks{ujjaldebnath@gmail.com}\\
Department of Mathematics, Indian Institute of Engineering\\
Science and Technology, Shibpur, Howrah-711 103, India.\\}

\date{}
\maketitle

\begin{abstract}
Here, we explore the dynamics of particle near the horizon of charged Mandal-Sengupta-Wadia (MSW) black hole in 2+1 dimensions. It results analysis of angular momentum and potential energy for null and time-like geodesics. We also appraise the high center-of-mass energy of coming particles from rest at infinity near the horizon of the charged MSW black hole in 2+1 dimension for the extremal case. Finally, we study the ISCO and MBCO radii for this type of black hole.
\end{abstract}

\section{Introduction:}

Recently, ``Two particle collision at the neighborhood of horizons of the black hole" becomes fascinated window in Astrophysics. This method, known as BSW effect, has been firstly achieved by Banados, Silk and West \cite{Banados} in 2009 for the Kerr black hole to get large center-of-mass (CM) energy if one of particles has critical angular momentum. To get this process, the black hole should be extremal. The CM energy for non-extremal black hole has been demonstrated by Jacobson and Sotiriou \cite{Jacobson} where they have pointed out the drawback of this process. So, there must exist astrophysical limitations for extremal Kerr black hole. Lake \cite{Lake} has found divergent CM energy at inner horizon for rotating Kerr black hole. Liu et al. \cite{Liu} have studied BSW effect of the Kerr-Taub-NUT space-time for extremal case. Grib and Pavlov \cite{Grib} have established disbursement of acceleration of the particles to produce high amount of CM energy for a rotating black hole in the case of extremal and non extremal both. Kerr-de Sitter black hole for non extremal case \cite{Li} acts as a particle accelerator with unbounded CM energy for two particle collisions. Wei et al. \cite{Wei} have studied the same thing for the Sen black hole. For non rotating black hole, Zaslavskill \cite{Zaslavskii} has studied particle acceleration. Extension of the BSW effect on particle acceleration on various type of black holes such as charged, non charged, rotating and non rotating black holes have been investigated in \cite{Zhu,Said,Abdujabbarow,Pradhan}.\\

Some attractions have been occurred on a new type of black hole, named as Regular black hole with non-singular curvature. Bardeen \cite{Bardeen} has introduced First Regular black hole. Then slowly many authors have worked on this type of various black holes in \cite{Ayon-Beato,Hayward,Elezalde}. Next Amir and Ghosh \cite{Amir} have studied the rotating Hayward black hole as particle accelerator. Harada \cite{Harada} has given an excellent review work on particle collision. Several authors have discussed particle collision in \cite{Zaslavskii1,Guo,Zhang}.\\

Now, the lower dimensional setting becomes more sharper than 3+1 dimensional case to get conceptual focus. To solve Einstein gravity equations, 2+1 dimensional black hole is exact solution than 3+1 dimensional black hole. Both 2+ 1 black hole and 3+ 1 black hole contribute same physical properties. Fortunately, Banados et al. \cite{Banados0} have brought 2+1 dimensional black hole firstly, named as Banados-Teitelboim-Zanelli (BTZ) black hole with negative cosmological constant which has opened a new window in classical, thermodynamical and quantum mechanics. Fernando \cite{Fernando,Fernando1} has investigated charged dilaton black hole as particle accelerator in 2+1 dimensions which plays a key role in string theory. Then another kind of 2+1 dimensional black hole has been studied by Mandal, Sengupta and Wadia \cite{Mandal}. Then spectroscopy, thermodynamics and quasinormal modes of MSW black hole in 2+1 dimensions have been discussed by Sebastian \cite{Sebastian,Sebastian1}.\\

Our motivation of this paper is to check the non-divergency or divergency (finite or infinite value) of the center-of-mass energy for both non-extremal and extremal cases of the charged MSW black hole in 2+1 dimensions to compare with BSW effect \cite{Banados,Fernando,Fernando1,Amir,Pradhan2,Pradhan3}. In this work, we study a brief history of the charged MSW black hole in 2+1 dimensions \cite{Mandal} and calculate the radii of outer and inner horizons of this black hole in section 2. In section 3, we investigate circular geodesics for time-like and null cases. For time-like geodesics, we calculate angular momentum and energy per unit mass. We study effective potential for both cases and impact parameter for null geodesics. Here, we also find out ISCO and MBCO radii for this type of black hole. In section 4, we calculate center-of-mass energy for particle acceleration near the outer/inner horizon of the charged MSW black hole in 2+1 dimensions. We also analyze center-of-mass energy for particle-photon collision in section 5 and photon-photon collision in section 6. Finally, we give an inference of our proposed work in section 7.\\

\section{A brief history of MSW Black hole:}

MSW black holes relate to the concept of federation of General Relativity and String theory. From \cite{Chan}, Einstein-Maxwell-dilaton action for (2+1) dimension is given by
\begin{equation}\label{1}
S=\int d^{3}x \sqrt{-g}\Big[R-\frac{B}{2}(\nabla\phi)^{2}-e^{-4a\phi}F_{\mu\nu}F^{\mu\nu}+2e^{b\phi}\Lambda\Big]
\end{equation}
where $a$, $b$, $\Lambda$ and $B$ are arbitrary couplings, $R$ is the Ricci scalar, $\phi$ is the dilaton field, $F_{\mu\nu}$ is Maxwell field. From \cite{Maki}, the corresponding action (1) for the low energy string theory with $B=8$,$b=4$,$a=1$ execute the conformal transformation
\begin{equation}\label{2}
g_{\mu\nu}^{S}=e^{\frac{4\phi}{(n-1)}}g_{\mu\nu}^{E}
\end{equation}
where $E$ and $S$ are Einstein and string metrics. Now we vary the aforementioned action (1) with metric, Maxwell and dilaton fields and we get
\begin{equation}\label{3}
R_{\mu\nu}=\frac{B}{2}\nabla_{\mu}\phi\nabla_{\nu}\phi+e^{-4a\phi}(-g_{\mu\nu}F^{2}+2F_{\mu}^{\alpha}F_{\nu\alpha})-2g_{\mu\nu}e^{b\phi}\Lambda,
\end{equation}
\begin{equation}\label{4}
\nabla^{\mu}(e^{-4a\phi}F_{\mu\nu})=0,
\end{equation}
\begin{equation}\label{5}
\frac{B}{2}(\nabla^{\mu}\nabla_{\mu}\phi)+2ae^{-4a\phi}F^{2}+be^{b\phi}\Lambda=0.
\end{equation}\\
Now for getting MSW 2+1 black hole we can obtain charged MSW 2+1 black hole solution with the help of equations (1)-(5) (taking $\phi=-\frac{1}{4}\ln(\frac{r}{\beta})$, $\beta$ is constant) given as \cite{Chan}
\begin{equation}\label{6}
ds^{2}=-H(r)dt^{2}+\frac{dr^{2}}{H(r)}+\beta rd\varphi^{2}
\end{equation}
where
\begin{equation}\label{7}
H(r)=8\Lambda\beta r-2m\sqrt{r}+8Q^{2}.
\end{equation}
Here, $\Lambda$ is cosmological constant, $m$ is the mass, $Q$ is the charge of the MSW black hole.\\

From where we can get two horizons: Event horizon and Cauchy horizon as $r_{\pm}=\frac{m^{2}-32\Lambda\beta Q^{2}\pm m\sqrt{m^{2}-64\Lambda\beta Q^{2}}}{32\Lambda^{2}\beta^{2}}$ when $m^{2}\geq 64\Lambda\beta Q^{2}$. Equality holds for extremal case with $r=\frac{m^{2}-32\Lambda\beta Q^{2}}{32\Lambda^{2}\beta^{2}}$.\\

Whenever $Q\rightarrow 0$ then $r_{+}=\frac{m^{2}}{16\Lambda^{2}\beta^{2}}$ and $r_{-}=0$ i.e., there exists only event horizon but no inner or cauchy horizon which corresponds to MSW 2+1 black hole with no charge.\\

\begin{figure}

$\hspace{4cm}$\includegraphics[height=2.0in]{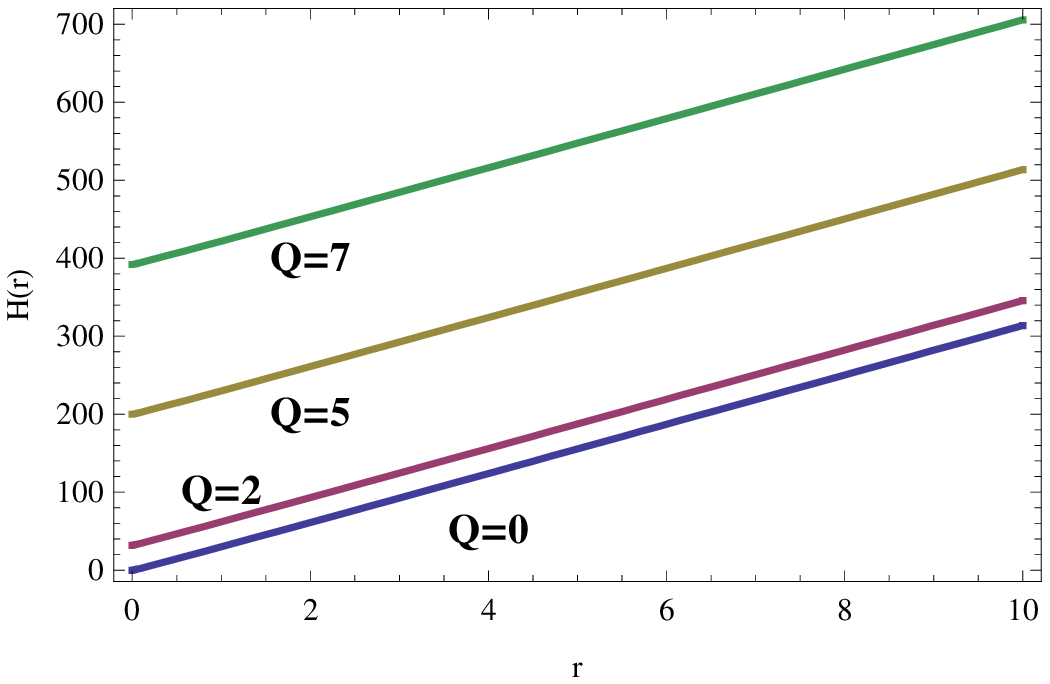}
\vspace{4 mm}
Fig.1\\
\vspace{4 mm}\\
Fig.1 represents the plots of $H(r)$ with respect to $r$ taking $m=1$, $\beta=2$ and $\Lambda=2$.
\end{figure}

\section{Circular Geodesics:}

Geodesics generalize the Euclidian straight line to curved space-time followed by a particle without any kind of force. The circular geodesics is defined by $r$=constant. In this section we do follow the geodesics equations for two circular orbits: Null geodesics and Time-like geodesics \cite{Chandrashekar}.\\

We consider the Lagrangian of the metric (6) as
\begin{equation}\label{8}
2\textit{L}=-H(r)\dot{t}^{2}+(H(r))^{-1}\dot{r}^{2}+\beta r \dot{\varphi}^{2}
\end{equation}
where the generalized momenta are given by
\begin{equation}\label{9}
\left.
\begin{array}{ll}
p_{t}\equiv \frac{\partial \textit{L}}{\partial \dot{t}}=-H(r)u^{t}\\
p_{r}\equiv \frac{\partial \textit{L}}{\partial \dot{r}}=\frac{u^{r}}{H(r)}\\
p_{\varphi}\equiv \frac{\partial \textit{L}}{\partial \dot{\varphi}}=\beta ru^{\varphi}
\end{array}
\right\}
\end{equation}
where $\textbf{u}=(u^{t},u^{r},u^{\varphi})$ is the three velocity vector in 2+1 dimensions. Now, we define the conserved energy $E$ and conserved angular momentum $L$ per unit mass respectively
\begin{equation}\label{10}
\left.
\begin{array}{ll}
u^{t}=\frac{E}{H(r)}\\
u^{\varphi}=\frac{L}{\beta r}
\end{array}
\right\}
\end{equation}\\

We take Hamiltonian, independent of $t$, being as
\begin{equation}\label{11}
2{\cal H}=-H(r)(u^{t})^{2}+(H(r))^{-1}(u^{r})^{2}+\beta r (u^{\varphi})^{2}=\sigma
\end{equation}
where $\sigma=0$ for light-like geodesics, $\sigma=-1$ for time-like geodesics and $\sigma=+1$ for space-like geodesics. Substituting the equation (10) in equation (11) we get the radial velocity as
\begin{equation}\label{12}
(u^{r})^{2}=E^{2}-H(r)\Big(\frac{L^{2}}{\beta r}-\sigma\Big)
\end{equation}
where the effective potential is as followed by
\begin{equation}\label{13}
\textit{V}_{eff}=\Big(\frac{L^{2}}{\beta r}-\sigma\Big)H(r).
\end{equation}\\

\subsection{Time-like Geodesics:}

With $\sigma=-1$, the radial equation (12) becomes
\begin{equation}\label{14}
(u^{r})^{2}=E^{2}-H(r)\Big(\frac{L^{2}}{\beta r}+1\Big)
\end{equation}
where
\begin{equation}\label{15}
\textit{V}_{eff}=\Big(\frac{L^{2}}{\beta r}+1\Big)H(r).
\end{equation}\\

We must have for circular geodesics

\begin{equation}\label{16}
\left.
\begin{array}{ll}
(u^{r})^{2}=0\\
\frac{d(u^{r})^{2}}{dr}=0
\end{array}
\right\}
\end{equation}\\

Now for the equations (7), (14), (15) and (16) we have angular momentum and energy per unit mass respectively
\begin{equation}\label{17}
L^{2}=\frac{\beta r^{\frac{3}{2}}(8\Lambda\beta\sqrt{r}-m)}{8Q^{2}-m\sqrt{r}}
\end{equation}
and
\begin{equation}\label{18}
E^{2}=\frac{4(4\Lambda\beta r-m\sqrt{r}+4Q^{2})^{2}}{8Q^{2}-m\sqrt{r}}.
\end{equation}\\

There exists circular motion for both angular momentum and energy being real and finite i.e.,
\begin{equation*}
8Q^{2}-m\sqrt{r}>0 \Rightarrow r<\frac{64Q^{4}}{m^{2}}.
\end{equation*}\\

We have plotted the graphs for radial components and effective potential with respect to radius for time-like geodesics. From figure 3 we have noticed that for the non charged MSW black hole in 2+1 dimensions the effective potential reaches zero at $r=0$, then increases with $r$. For $Q\neq 0$ at first the effective potential $\textit{V}_{eff}\rightarrow \infty$ for $r\rightarrow 0$ then decreases very fast from infinity with $r$ but after a certain value of $r$ the effective potential increases slowly comparable to the former case with radius. In this case the effective potential never reach zero.\\

\subsection{ISCO and MBCO:}

Kaplan \cite{1KSA49} has first observed a circular orbit which is innermost stable for the Schwarzschild black hole with a minimal radius $3r_{h}$ ($r_{h}$ is the radius of the horizon of the Schwarzschild black hole). This orbit is known as Innermost Stable Circular Orbit (ISCO) at which a test particle can stably orbit a massive object. Innermost circular orbit has been investigated by Jefremov et al. \cite{1JPITOYKGSB15} for Schwarzschild and Kerr metrices. Zhang et al. \cite{1ZYPWSWGWDSTTLYX18} has also studied the ISCO for the Kerr-Newman black hole with spinning test particle. The ISCO in different black holes have been investigated in refs. \cite{1ZJLHPRDGGE00,1MAGPSWM04,1SPFSB07,1HSKBBECV11,1ASBLDTSN12,1HS13,1CC14,1HS14,1ZOB15,1DTRJVSR15}. According to \cite{2BJMPWHTSA72}, all spherical orbits are not bounded of the black hole. On an infinitesimal perturbation, a particle can escape to infinity from the unbounded geodesics. So, there exists a critical orbit, known as Marginally Bound Circular Orbit (MBCO), which is separated from the unbounded spherical orbits. Hod \cite{2HS13} has studied MBCO for the rotating black holes.\\

We get ISCO radius ($r_{ISCO}$) of this type of black hole with equation (16) and $\frac{d^{2}(u^{r})^{2}}{dr^{2}}=0$ \cite{Pradhan2} given by ISCO equation
\begin{equation}\label{19}
144m^{2}\Lambda^{2}\beta^{2}r^{2}-r(m^{4}-32m^{2}Q^{2}\Lambda\beta+16384\Lambda^{2}\beta^{2}Q^{4})+144m^{2}Q^{4}=0.
\end{equation}\\

MBCO radius ($r_{MBCO}$) \cite{Pradhan2} is given by the MBCO equation
\begin{equation}\label{20}
m^{2}r(1-32\Lambda\beta r-32Q^{2})^{2}=16(2Q^{2}-16Q^{4}-32\Lambda\beta Q^{2}r-16\Lambda^{2}\beta^{2}r^{2})^{2}.
\end{equation}

If $r>r_{ISCO}$ then the circular orbits are stable and as well as if $r>r_{MBCO}$ then we can conclude that the spherical orbits are bounded for the charged MSW black hole in 2+1 dimensions. Beyond theses conditions we can not get stable as well as bounded circular orbit. In astrophysics, these orbits play a crucial role to give important information of the background geometry.\\

When $Q\rightarrow0$ then $r_{ISCO}=\frac{144\Lambda^{2}\beta^{2}}{m^{2}}$ and $r_{MBCO}$ can be found by solving the equation $m^{2}r(1-32\Lambda\beta r)^{2}=16(m^{2}r+16\Lambda^{2}\beta^{2}r^{2})^{2}$ respectively which correspond to MSW 2+1 black hole with no charge.\\

\begin{figure}

\includegraphics[height=2.0in]{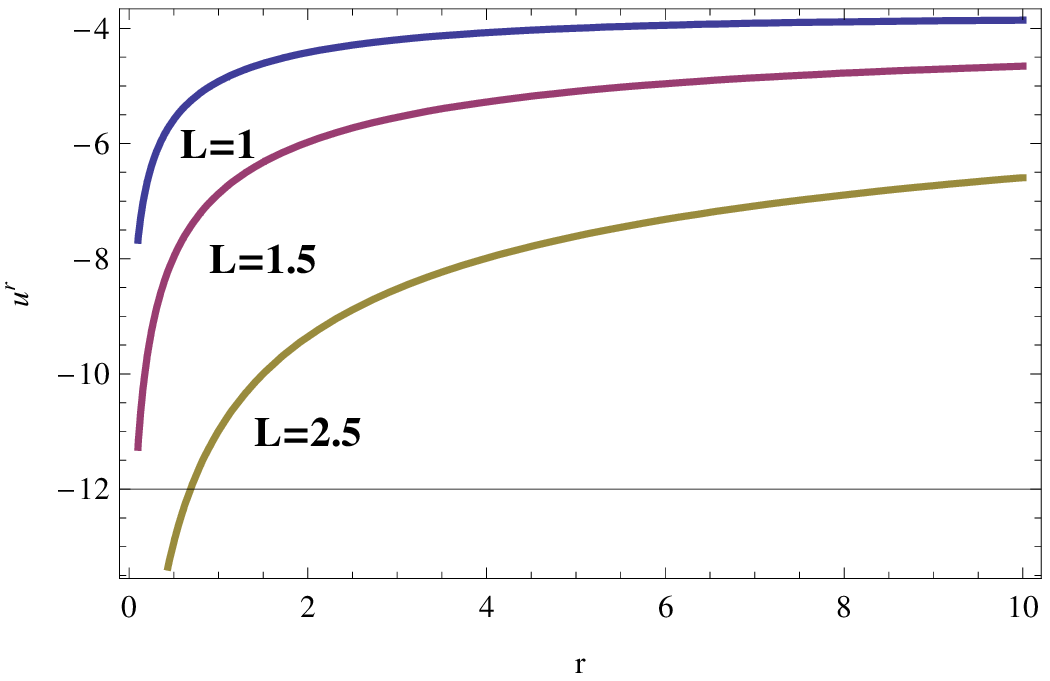}~~~
\includegraphics[height=2.0in]{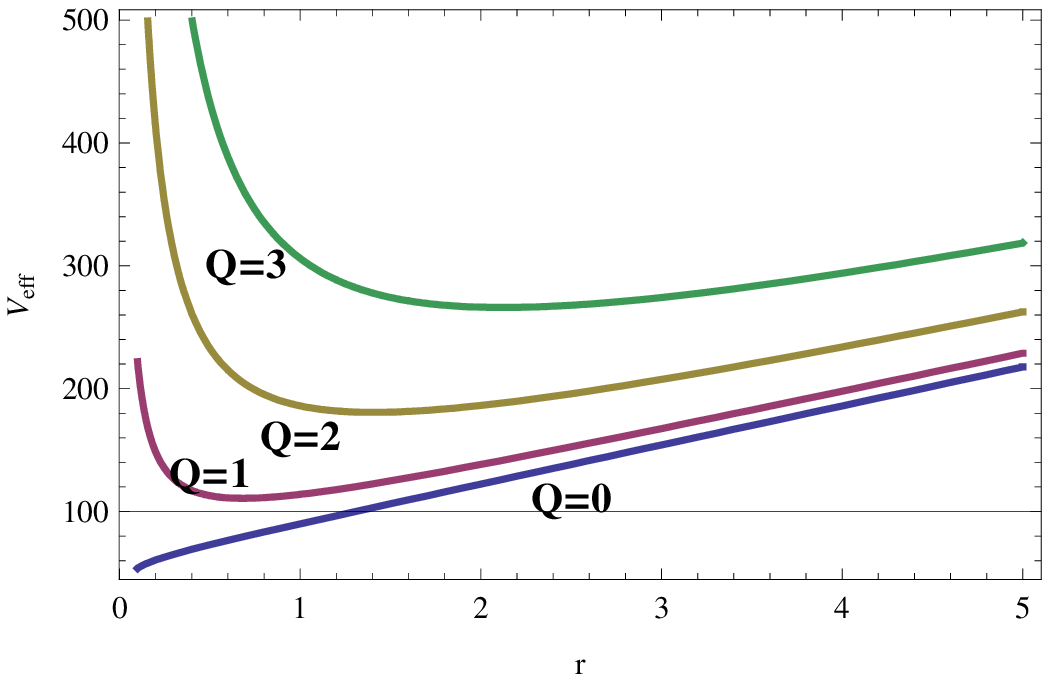}
\vspace{4 mm}
~~~~~~~~~~~~~~~~~~~~~~~~~~~~Fig.2 ~~~~~~~~~~~~~~~~~~~~~~~~~~~~~~~~~~~~~~~~~~~~~~~~~~~~~~~~~~~~~~~~~~Fig.3 \\
\vspace{4 mm}\\
Figs. 2 and 3 represent the plots of the radial component $u^{r}$ and the effective potential $V_{eff}$ for time-like geodesics with respect to $r$ taking $m=1$, $\beta=0.1$ and $\Lambda=0.1$.
\end{figure}

\subsection{Light rays orbit:}

Now for this orbit having $\sigma=0$, equation (12) reduces to
\begin{equation}\label{21}
U_{eff}=L^{2}\Bigg[8\Lambda-\frac{2m\sqrt{r}-8Q^{2}}{\beta r}\Bigg].
\end{equation}\\

For light rays/photon orbit we take
\begin{equation}\label{22}
\left.
\begin{array}{ll}
U_{eff}=E^{2}\\
\frac{dU_{eff}}{dr}=0
\end{array}
\right\}
\end{equation}\\

Substituting equation (21) in the above condition (22) we have
\begin{equation}\label{23}
\frac{E}{L}=\pm\sqrt{8\Lambda-\frac{2m\sqrt{r}-8Q^{2}}{\beta r}}
\end{equation}
and $r=\frac{64Q^{4}}{m^{2}}$ corresponding to the circular photon orbital radius of this black hole. When $Q\rightarrow0$ then there exists no radius for circular photon orbit i.e., the non-charged MSW black hole in 2+1 dimensions has no null geodesics.\\

Now, we define the impact parameter \cite{Pradhan2,1DTK12} which determines the shape of the orbit of a photon incident from infinity on a black hole given by
\begin{equation}\label{24}
D_{IF}=+\frac{L}{E}=\frac{1}{\sqrt{8\Lambda-\frac{2m\sqrt{r}-8Q^{2}}{\beta r}}}
\end{equation}
whereas $D_{IF}=\frac{1}{\sqrt{8\Lambda-\frac{2m}{\beta\sqrt{r}}}}$ for the non-charged MSW black hole in 2+1 dimensions.\\

We have plotted the graphs for effective potential with respect to radius for light rays orbit. From figure 4 we have noticed that for charged MSW black hole in 2+1 dimensions the effective potential never reach zero. But for non charged ($Q=0$) case the effective potential $\textit{U}_{eff}\rightarrow -\infty$ as $r\rightarrow 0$ that means the effective potential reaches zero from negative infinity then increases with $r$. Again for $Q\neq 0$ the effective potential decreases rapidly from positive infinity to finite value with $r$ but after some certain value of $r$ the effective potential decreases extreme slowly.\\

\begin{figure}

$\hspace{4cm}$\includegraphics[height=2.0in]{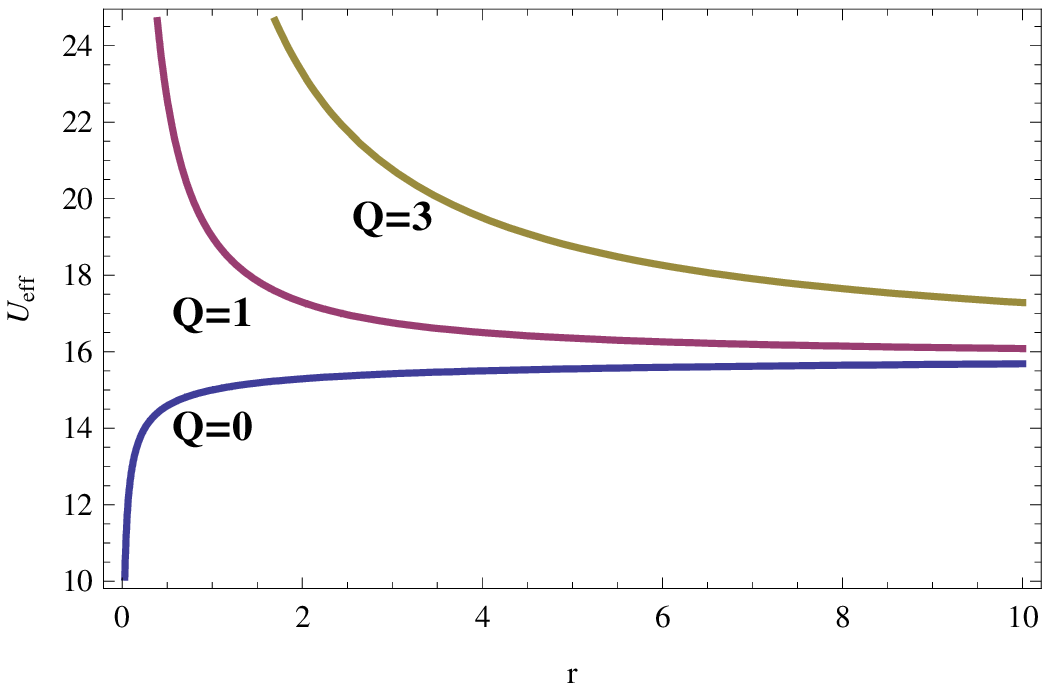}
\vspace{4 mm}
Fig.4 ~~~~~~~~~~~~~~~~~~~~~~~~~~~~~~~~~~~~~ \\
\vspace{4 mm}\\
Figs. 4 represents the plots of the effective potential $U_{eff}$ for null geodesics with respect to $r$ taking $m=0.1$, $\beta=0.2$ and $\Lambda=0.1$.
\end{figure}

\section{CME for particle collision near the horizon of MSW black hole in 2+1 dimensions:}

Inspired by BSW method \cite{Banados} we do the calculation of center-of-mass energy for time-like geodesical particle collision for this type of black hole for which three velocity components are given by
\begin{equation}\label{25}
\left.
\begin{array}{ll}
u^{t}=\dot{t}=\frac{E}{H(r)}\\
u^{r}=\dot{r}=\pm\sqrt{E^{2}-H(r)(1+\frac{L^{2}}{\beta r})}\\
u^{\varphi}=\dot{\varphi}=\frac{L}{\beta r}
\end{array}
\right\}
\end{equation}
provided $E^{2}>H(r)(1+\frac{L^{2}}{\beta r})$ for infall/escape of the particles. In equation (25), if $u^{r}>0$ then there will be outgoing geodesics and if $u^{r}<0$ then it is seen incoming geodesics for the particles. Here, we take ``$-$" sign of $u^{r}$ for the particles falling into the MSW black hole in 2+1 dimensions. So, the corresponding three velocity components (25) reduce to
\begin{equation}\label{26}
u_{i}^{a}=\Bigg(\frac{E_{i}}{H(r)},-Z_{i},\frac{L_{i}}{\beta r}\Bigg)
\end{equation}
where
\begin{equation}\label{27}
Z_{i}=\sqrt{E_{i}^{2}-H(r)(1+\frac{L_{i}^{2}}{\beta r})}.
\end{equation}\\

\subsection{Two neutral particles with different mass:}

We study CM energy for two colliding particles with rest masses $M_{1}$, $M_{2}$; angular momenta $L_{1}$, $L_{2}$ per unit mass and energy $E_{1}$, $E_{2}$ per unit mass. Therefore, we consider that two particles are at the same space-time point having the three momenta \cite{Liu}
\begin{equation*}
p_{i}^{a}=M_{i}u_{i}^{a}
\end{equation*}
where $p_{i}^{a}$ and $u_{i}^{a}$ are the three momenta and the three velocity of particles $i$ ($i = 1, 2$). Now, the sum of these two momenta is
\begin{equation*}
p_{t}^{a}=p_{1}^{a}+p_{2}^{a}.
\end{equation*}\\

Then the CM energy for the two particles is defined as
\begin{equation}\label{28}
E_{CM}^{2}=-p_{t}^{a}p_{ta}=-(M_{1}u_{1}^{a}+M_{2}u_{2}^{a})(M_{1}u_{1a}+M_{2}u_{2a}).
\end{equation}\\

Using $u^{a}u_{a}=-1$, we have \cite{Banados,Li,Wei,Zaslavskii,Zhu,Said}
\begin{equation}\label{29}
E_{CM}^{2}=2M_{1}M_{2}\Bigg[\frac{(M_{1}-M_{2})^{2}}{2M_{1}M_{2}}+(1-g_{\lambda\mu}u_{1}^{\lambda}u_{2}^{\mu})\Bigg].
\end{equation}\\

Using equations (25), (26), (27)and (29) we get the CM energy as
\begin{equation}\label{30}
E_{CM}^{2}=2M_{1}M_{2}\Bigg[\frac{(M_{1}-M_{2})^{2}}{2M_{1}M_{2}}+\Bigg(1+\frac{E_{1}E_{2}}{H(r)}-\frac{Z_{1}Z_{2}}{H(r)}-\frac{L_{1}L_{2}}{\beta r}\Bigg)\Bigg].
\end{equation}\\

\begin{figure}

\includegraphics[height=2.0in]{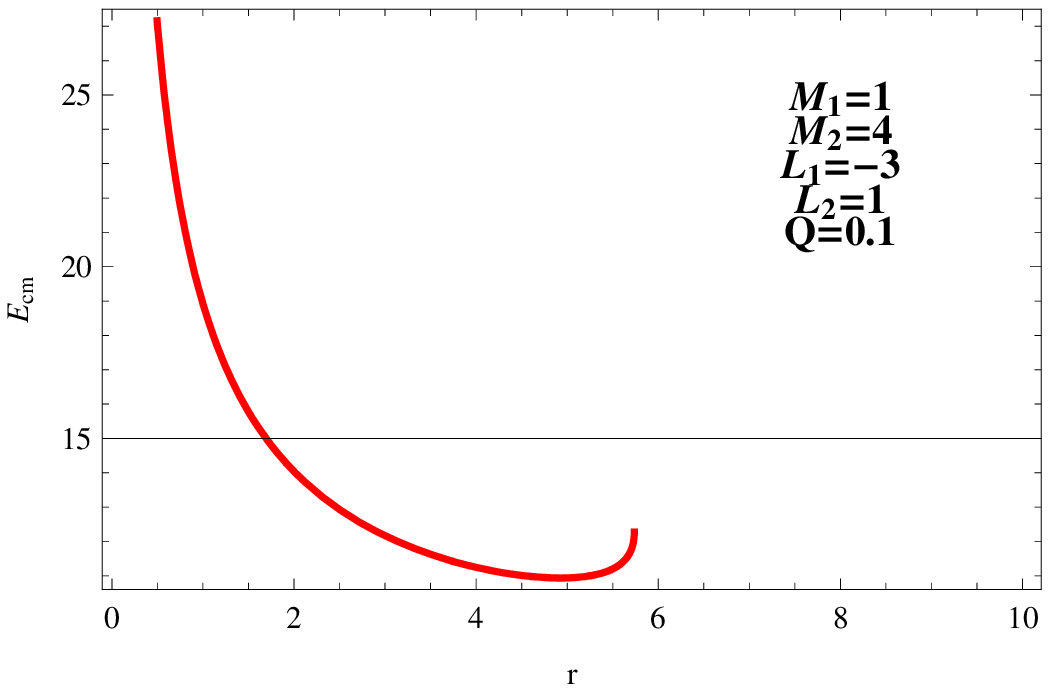}~~~
\includegraphics[height=2.0in]{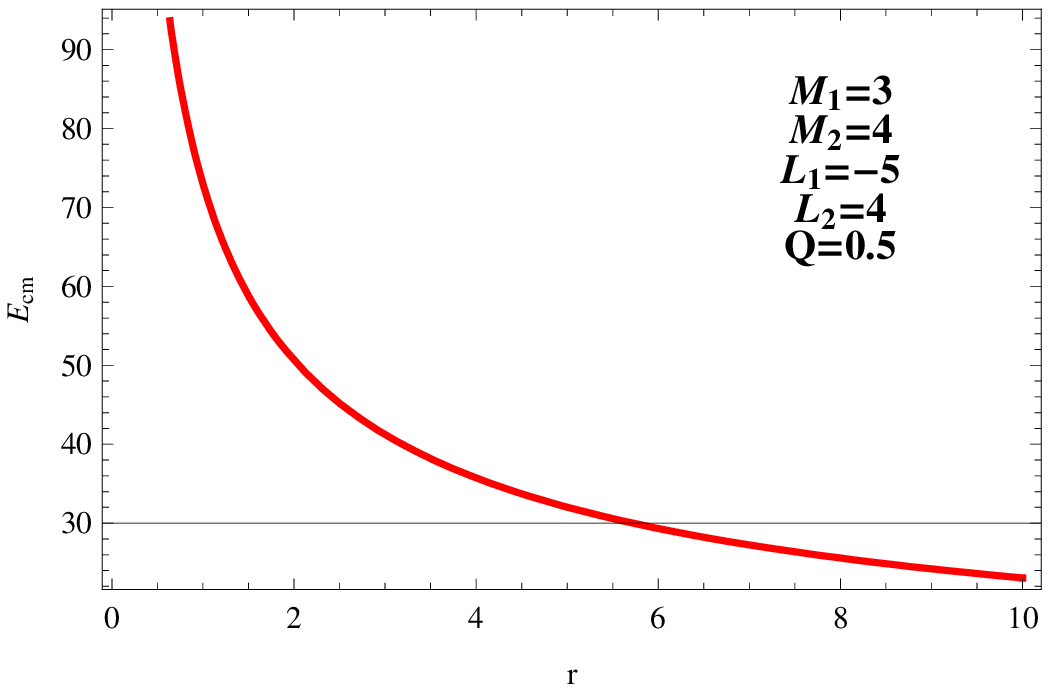}
\vspace{4 mm}
~~~~~~~~~~~~~~~~~~~~~~~~~~~~Fig.5 ~~~~~~~~~~~~~~~~~~~~~~~~~~~~~~~~~~~~~~~~~~~~~~~~~~~~~~~~~~~~~~~~~~Fig.6 \\
\vspace{4 mm}\\
Figs. 5 and 6 represent the plots of $E_{CM}$ with respect to $r$ taking $m=1$, $\beta=5$ and $\Lambda=2$.
\end{figure}

\subsection{Two neutral particles of same masses:}
With the help of previous section for colliding particles with rest mass $M$, we define CM energy formula as
\begin{equation}\label{31}
\Bigg(\frac{E_{CM}}{\sqrt{2}M}\Bigg)^{2}=1+\frac{E_{1}E_{2}}{H(r)}-\frac{Z_{1}Z_{2}}{H(r)}-\frac{L_{1}L_{2}}{\beta r}.
\end{equation}

For non-extremal case i.e.,when $r_{+}\neq r_{-}$, then CM energy (31) near the event horizon reduces to
\begin{equation}\label{32}
E_{CM}|_{r=r_{+}}=\sqrt{2}M\sqrt{1-\frac{L_{1}L_{2}}{\beta r_{+}}}.
\end{equation}
It is clear from equation (32) that for $r_{+}\neq r_{-}$ (non-extremal case), the CM energy is finite which depends upon the values of the angular momenta. Now, we draw the graphs for $E_{CM}$ with respect to $r$ and also locate $r_{+}=6.15$ on the figure 7 and $r_{+}=4.51$ on the figure 8. \\

The angular velocity at $r=r_{+}$ is given as
\begin{equation}\label{33}
\Omega_{h}=\frac{\sqrt{8\Lambda\beta\sqrt{r}-m}}{\beta^{\frac{1}{2}}r^{\frac{1}{4}}}.
\end{equation}

Then the critical angular momenta is
\begin{equation}\label{34}
L_{i}=\frac{E_{i}}{\Omega_{h}}.
\end{equation}
For extremal case, when $m^{2}=64\Lambda\beta Q^{2}$ then the horizon is at $r=r_{+}=r_{-}=\frac{m^{2}}{64\Lambda^{2}\beta^{2}}$ and if one of particles has divergent angular momentum at the horizon i.e., $L_{1}\rightarrow \infty$ as $r \rightarrow \frac{m^{2}}{64\Lambda^{2}\beta^{2}}$, then we get $E_{CM}\rightarrow \infty$. So, if one of particles has very high angular velocity at the time of coincidence of the event horizon and cauchy horizon on each other ($r_{+}=r_{-}$) then we have extremely high center-of-mass energy.\\

At the center of this MSW black hole, the center-of-mass energy is also divergent ($E_{CM}\rightarrow \infty$). When $r\rightarrow \infty$ then $E_{CM}\rightarrow \sqrt{2}M$.\\

\begin{figure}

\includegraphics[height=2.0in]{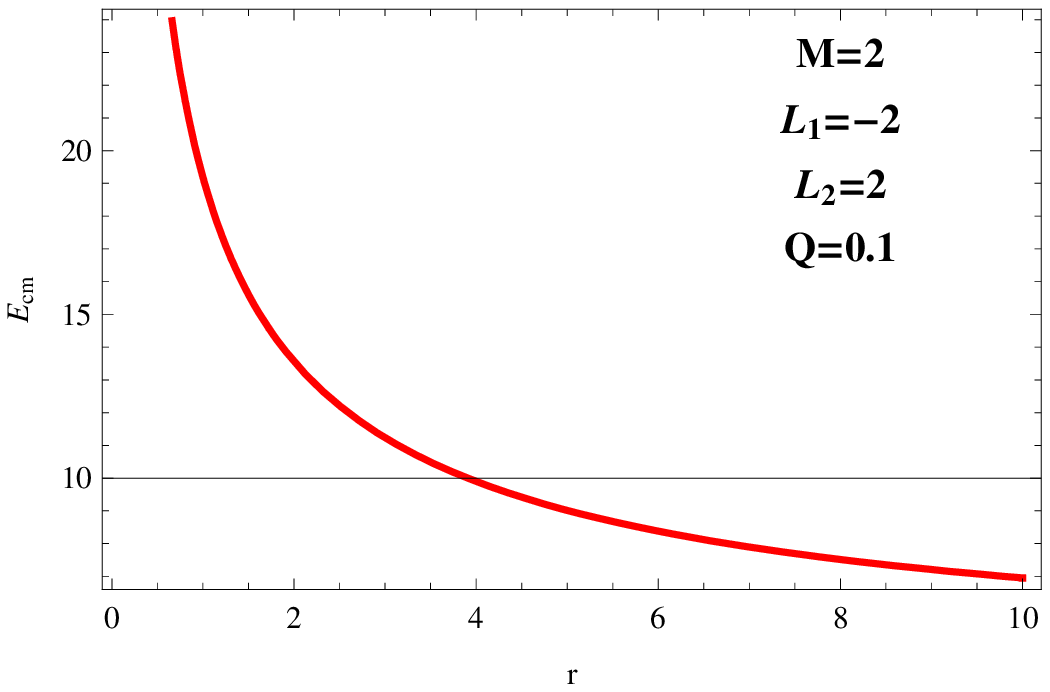}~~~
\includegraphics[height=2.0in]{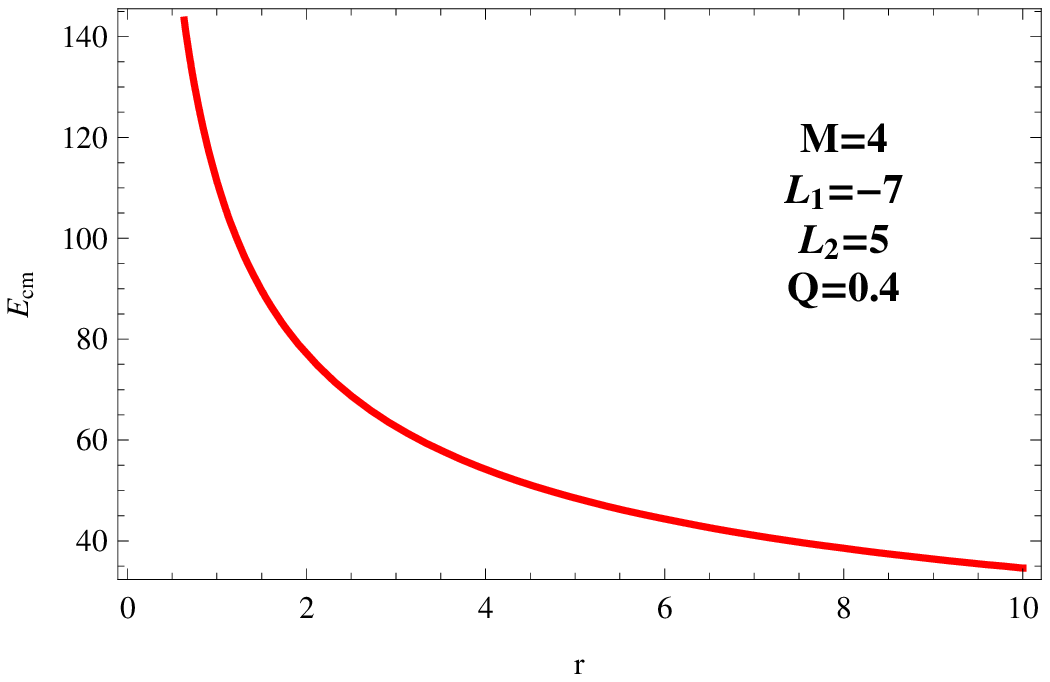}
\vspace{4 mm}
~~~~~~~~~~~~~~~~~~~~~~~~~~~~Fig.7 ~~~~~~~~~~~~~~~~~~~~~~~~~~~~~~~~~~~~~~~~~~~~~~~~~~~~~~~~~~~~~~~~~~Fig.8 \\
\vspace{4 mm}\\
Figs. 7 and 8 represent the plots of $E_{CM}$ with respect to $r$ taking $m=2$, $\beta=2$ and $\Lambda=0.1$.
\end{figure}

\section{Particle collision with photon:}
Scattering of a photon by an electron incorporates Compton Scattering process where the energy of the photon is transferred to the recoiled electron. Motivated by this method we consider an infalling particle collision with an outgoing massless photon\cite{Halilsoy}. With the help of time-like geodesic of a particle and the null geodesic of a photon i.e., with the help of equations
\begin{equation*}
g_{\mu\nu}U^{\mu}U^{\nu}=-1
\end{equation*}
and
\begin{equation*}
g_{\mu\nu}K^{\mu}K^{\nu}=0
\end{equation*}
respectively, we get photon velocity components ($\sigma=0$) as
\begin{equation}\label{35}
K^{t}=\frac{E_{\gamma}}{H(r)},
\end{equation}
\begin{equation}\label{36}
K^{\varphi}=\frac{L_{\gamma}}{\beta r},
\end{equation}
\begin{equation}\label{37}
(K^{r})^{2}=E_{\gamma}^{2}-\frac{L_{\gamma}^{2}}{\beta r}H(r)
\end{equation}
where $(U^{t}, U^{r}, U^{\varphi})$ and $(K^{t}, K^{r}, K^{\varphi})$ are particle velocity components and photon velocity components respectively, $H(r)$ is given from equation (7) and $E_{\gamma}$, $L_{\gamma}$ are energy and angular momentum of the photon respectively.\\

The center-of-mass for collision between a particle of rest mass $M$ and a mass free photon is given as \cite{Halilsoy}
\begin{equation}\label{38}
E_{CM}^{2}=M^{2}-2Mg_{\mu\nu}U^{\mu}K^{\nu}.
\end{equation}
Substituting equations (25), (35), (36) and (37) in equation (38), we obtain
\begin{equation}\label{39}
E_{CM}^{2}=M^{2}-2M\Bigg[-\frac{EE_{\gamma}}{H(r)}+\frac{\sqrt{E^{2}-H(r)\Bigg(1+\frac{L^{2}}{\beta r}\Bigg)}\sqrt{E_{\gamma}^{2}-\frac{L_{\gamma}^{2}}{\beta r}H(r)}}{H(r)}+\frac{LL_{\gamma}}{\beta r}\Bigg].
\end{equation}\\

\begin{figure}

\includegraphics[height=2.0in]{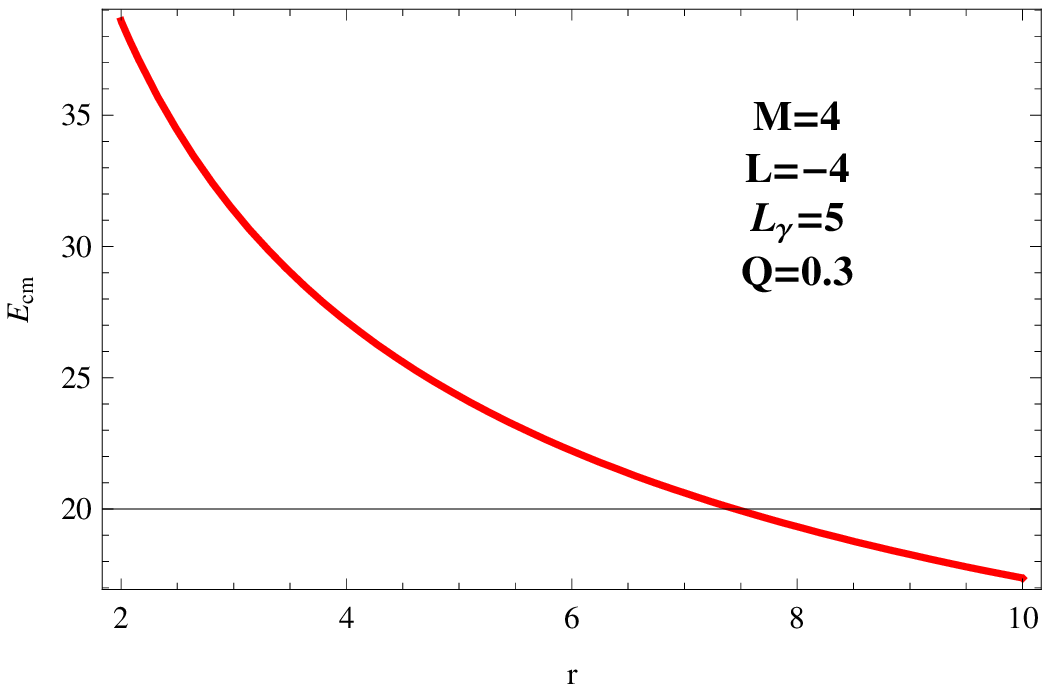}~~~
\includegraphics[height=2.0in]{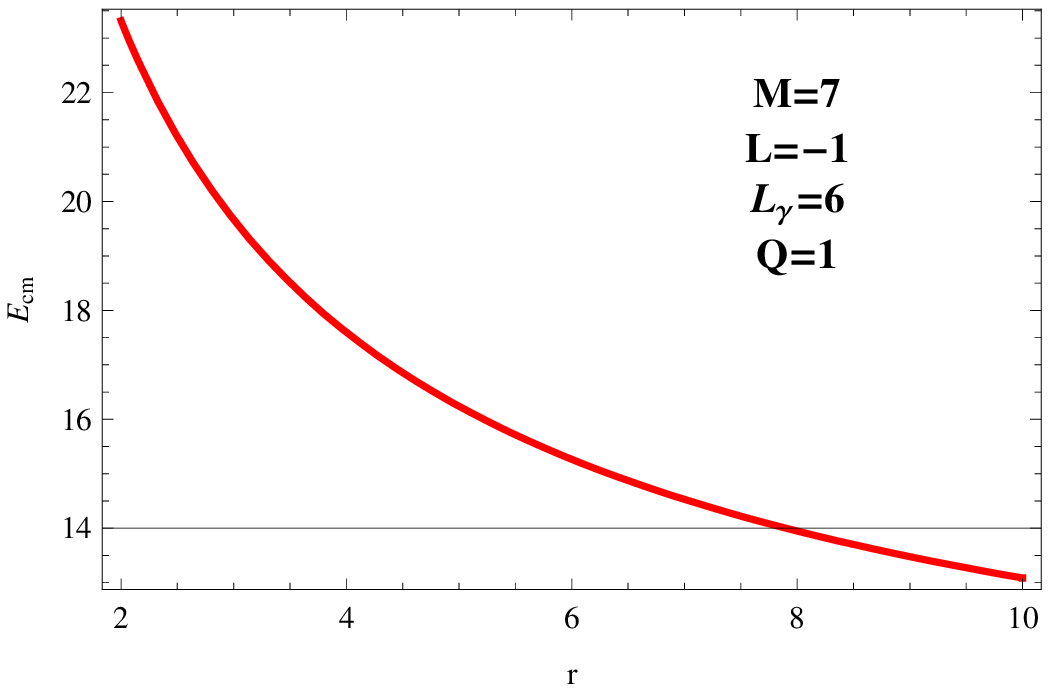}
\vspace{4 mm}
~~~~~~~~~~~~~~~~~~~~~~~~~~~~Fig.9 ~~~~~~~~~~~~~~~~~~~~~~~~~~~~~~~~~~~~~~~~~~~~~~~~~~~~~~~~~~~~~~~~~~Fig.10 \\
\vspace{4 mm}\\
Figs. 9 and 10 represent the plots of $E_{CM}$ with respect to $r$ taking $m=0.1$, $\beta=0.1$ and $\Lambda=0.2$.
\end{figure}

\section{Collision between two photons:}
Here we consider collision of two massless photons near the horizon \cite{Halilsoy} of this black hole. The null geodesic of two photons satisfying
\begin{equation*}
g_{\mu\nu}K^{\mu}K^{\nu}=0,
\end{equation*}
the photon velocity components ($\sigma=0$) are
\begin{equation}\label{40}
K_{i}^{t}=\frac{E_{\gamma_{i}}}{H(r)},
\end{equation}
\begin{equation}\label{41}
K_{i}^{\varphi}=\frac{L_{\gamma_{i}}}{\beta r},
\end{equation}
\begin{equation}\label{42}
(K_{i}^{r})^{2}=E_{\gamma_{i}}^{2}-\frac{L_{\gamma_{i}}^{2}}{\beta r}H(r)
\end{equation}
where $H(r)$ is given by equation (7) and $E_{\gamma_{i}}$, $L_{\gamma_{i}}$ are energy and angular momentum of the $i^{th}$ photon respectively, $i=1,2$. Now the center-of-mass energy for collision of two mass free photons is \cite{Halilsoy}
\begin{equation}\label{43}
E_{CM}^{2}=-2g_{\mu\nu}K_{1}^{\mu}K_{2}^{\nu}.
\end{equation}
Now equation (43) can be reduced with the help of equations (40), (41), and (42) as
\begin{equation}\label{44}
E_{CM}^{2}=-2\Bigg[-\frac{E_{\gamma_{1}}E_{\gamma_{2}}}{H(r)}+\frac{\sqrt{E_{\gamma_{1}}^{2}-\frac{L_{\gamma_{1}}^{2}}{\beta r}H(r)}\sqrt{E_{\gamma_{2}}^{2}-\frac{L_{\gamma_{2}}^{2}}{\beta r}H(r)}}{H(r)}+\frac{L_{\gamma_{1}}L_{\gamma_{2}}}{\beta r}\Bigg].
\end{equation}\\

\begin{figure}

\includegraphics[height=2.0in]{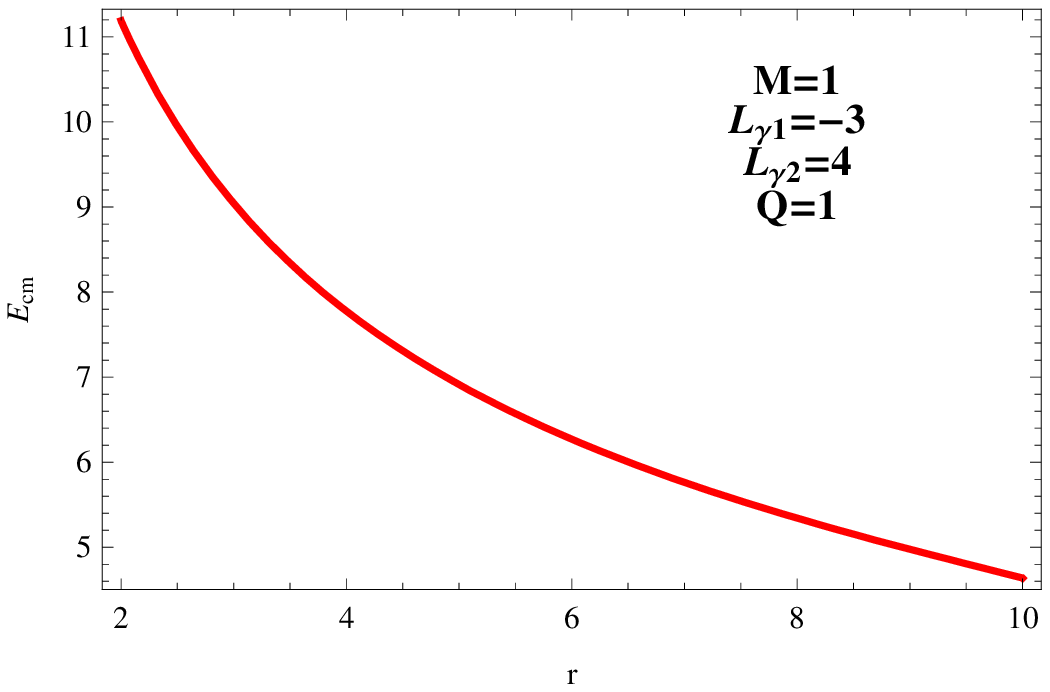}~~~
\includegraphics[height=2.0in]{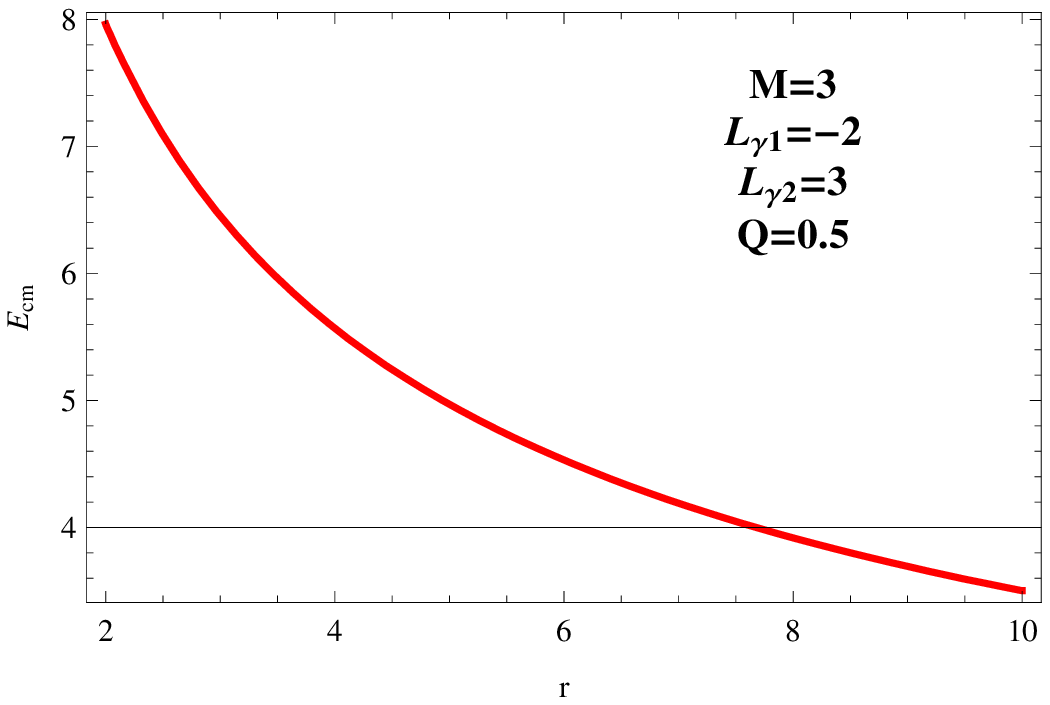}
\vspace{4 mm}
~~~~~~~~~~~~~~~~~~~~~~~~~~~~Fig.11 ~~~~~~~~~~~~~~~~~~~~~~~~~~~~~~~~~~~~~~~~~~~~~~~~~~~~~~~~~~~~~~~~~~Fig.12 \\
\vspace{4 mm}\\
Figs. 11 and 12 represent the plots of $E_{CM}$ with respect to $r$ taking $m=1$, $\beta=0.1$ and $\Lambda=0.2$.
\end{figure}

\section{Inference:}

In this work, a charged MSW black hole has been observed in 2+1 dimensions. From this type of black hole we have calculated cauchy and event horizon radii whereas there is no cauchy/inner horizon in case of this black hole with no charge.\\

Next, we have considered two geodesics: (i) Time-like geodesics and (ii) Light-ray geodesics. For time-like geodesics we have calculated three velocity components \textbf{i.e.,} $u^{t}$, $u^{r}$ and $u^{\varphi}$. Also from figure 3, we have found effective potential which reaches zero at $r=0$, then increases with $r$. For $Q\neq 0$ the effective potential $\textit{V}_{eff}\rightarrow \infty$ for $r\rightarrow 0$ but after a certain value of $r$, it increases slowly with radius. We have solved energy and angular momentum per unit mass of this black hole; we have also found ISCO and MBCO radii for time-like geodesics from where we have concluded that the circular orbits are stable for $r>r_{ISCO}$ and bounded for $r>r_{MBCO}$. For light rays orbit we have investigated the same things like time-like geodesics. Here, we have seen from figure 4 that for charged MSW black hole in 2+1 dimensions the effective potential never reach zero whereas for non charged ($Q=0$) case the effective potential ($U_{eff}$) crosses zero and tends to $-\infty$ as $r\rightarrow 0$. Again, for charged case the effective potential decreases from positive infinity to give finite value with $r$.\\

There has been drawn attention on the non-divergency and divergency of CME for particle collision near many black holes in \cite{Amir,Pradhan2,Pradhan3}. In this work, we have examined CME near the horizons of this black hole with particle collision to compare with the previous work to verify what we have got the values of CME (finite or infinite). Finally, we have noticed that $E_{CM}\rightarrow \infty$ for extremal case and finite for non-extremal case with same mass particle collision like the previous literatures \cite{Amir,Pradhan2,Pradhan3}. The CME is divergent near the center of black hole and $E_{CM}\rightarrow \sqrt{2}M$ when $r\rightarrow \infty$. So, for the non-extremal case and at infinity we have got non-divergent center-of-mass energy ($E_{CM}$ is finite) whereas for extremal case and at the center of this black hole it has been seen divergency of center-of-mass energy ($E_{CM}\rightarrow \infty$). Next, we have followed compton scattering process to obtain center-of-mass energy for particle-photon collision and photon-photon collision where photon is coming from Hawking radiation in presence of charged MSW black hole in 2+1 dimensions.\\

However, in this work we have checked whether there is any possibility to get high CME for the particle collision near MSW black hole in 2+1 dimensions like spinning dilaton black hole in 2+1 dimensions \cite{Fernando,Fernando1}. Finally, we have noticed infinite CME if two particles collide with this type of black hole in 2+1 dimensions. So, this work would be interesting to study MSW black hole in 2+1 dimensions if the particles could generate high center-of-mass energy.\\

\end{document}